\documentclass[10pt,a4papper]{article}
\usepackage[utf8]{inputenc}
\usepackage{indentfirst}
\usepackage[]{graphicx}
\usepackage{geometry} 
\usepackage{tikz-feynman}
\usepackage{textcomp}
\usepackage{subcaption}
\usepackage{cite}
\begin{document}
	\title{\textbf{Decay widths of radially excited vector mesons of the $U(3)$ nonet with production of three pseudoscalar mesons in the extended NJL model}}
	\author{M. K. Volkov$^{1}$\footnote{volkov@theor.jinr.ru}, A. A. Pivovarov$^{1}$\footnote{pivovarov@theor.jinr.ru} , K. Nurlan$^{1, 2}$\footnote{nurlan@theor.jinr.ru}\\
		\small
		\emph{$^{1}$Bogoliubov Laboratory of Theoretical Physics, JINR, 
                 141980 Dubna, Moscow region, Russia}\\	
		\small
		\emph{$^{2}$The Institute of Nuclear Physics, Almaty, 050032, Kazakhstan}}
	\date{}
	\maketitle
	\small
	
\begin{abstract}
In the extended NJL model, the decays of the first radially excited vector mesons $\rho'$,$\omega'$, $\phi'$ and ${K^*}'$ with production of $3P$, where $P=K,\pi,\eta$ are calculated. The box channels and the channels with intermediate vector mesons in the ground and first radially excited states are considered. The obtained results can be considered as predictions for future experiments.
\end{abstract}

\large

\section{\label{Intro}Introduction}
Research of the meson structure and mechanisms of their interactions in the nonperturbative area of the Quantum Chromodynamics (QCD) is one of the important tasks in physics of elementary particles. The first radially excited states of the vector mesons $\rho’$, $\omega’$, $\phi’$ and $K^{*’}$ contain unique information about the inner structure and dynamics of meson interaction at low energies~\cite{ParticleDataGroup:2024cfk}. Experimental research into processes of electron-positron annihilation into hadrons gives data on the main characteristics of excited mesons such as mass and width, and decay channels. This research is supplemented by measurements of hadron decays of the $\tau$ lepton because the mechanisms of the meson production are similar and can be explained in the framework of the unified model. At the same time, the $\tau$ lepton restricts the energy of the produced mesons with its mass, and imposes additional restrictions on masses of the mesons making contribution to intermediate channels. Indeed, taking into account the channels with intermediate mesons in the ground and first radially excited states turns out to be sufficient to describe meson $\tau$ decays~\cite{Volkov:2017arr}.

Among various processes containing excited mesons the decays of vector mesons into three pseudoscalar states are of special interest. In the energy range up to 2~GeV, the processes $e^+e^- \to 3P$ with intermediate channels $V, V’ \to 3P$ are important while calculating the cross section of the process, which is important for calculation of the anomalous magnetic moment of the muon. The experimental data accumulated by the collaborations SND, CMD-3 at the collider VEPP-2000, and also by the collaborations BaBar at PEP-II, Belle at KEK, and BES-III at BEPC II at higher energies allow one to study in detail the spectroscopy of excited mesons and various decay channels of these states. Particularly, based on the analysis of the cross section of the processes $e^+e^- \to\pi\pi\eta$ and $e^+e^- \to\pi\pi\pi$, the experimental values of the masses and widths of the mesons $\rho(1450)$ and $\omega(1420)$ were obtained~\cite{Achasov:2003ir,SND:2014rfi,Aulchenko:2015mwt,BaBar:2018erh,Achasov:2024tfh,Dimova:2025dxn}. The parameters of the first excited meson $\phi(1680)$ were fitted from the data $e^+e^- \to KK\pi$ and $e^+e^- \to KK\eta$, where the contribution of the channel with the decay $\phi' \to K^*K$ is dominant in the process $e^+e^- \to KK\pi$ and the decay $\phi'\to\phi\eta$ in the process with production of $KK\eta$~\cite{Ivanov:2019crp,SND:2020qmb, BESIII:2022wxz}. However, in these experiments, the individual widths of the decay channels of excited vector mesons into final states were not measured.

In addition to their importance for determining the cross section of $e^+e^-$ annihilation into hadrons, the decays $V,V’ \to 3P$ turn out to be intermediate channels in the weak and electromagnetic decays of heavy hadrons. In the decays of neutral and charged $B$ mesons, the channels with the production of virtual radially excited vector mesons $\rho’$, $\omega’$, $\phi’$ are observed~\cite{BaBar:2009vfr,Belle:2022dgi}. The channels with the excited vector mesons $\rho’$ and $K^{*’}$ are also important for the description of the decays of charmoniums and charm mesons into hadrons~\cite{FOCUS:2009bwp,BESIII:2014oag,BaBar:2017dwm,KEDR:2022dhm}.

From a theoretical point of view, the description of strong interactions in the low energy area faces fundamental difficulties related to inapplicability of the QCD Perturbation theory due to the large value of the coupling constant. Thus, various phenomenological models based on the fundamental principles of QCD, first of all chiral symmetry and the mechanism of its spontaneous breaking, as well as the hypothesis of vector dominance, are widespread. The Nambu--Jona-Lasinio (NJL) model is among these models. It is a phenomenological model based on the effective chirally symmetric four-quark interaction~\cite{Nambu:1961tp,Eguchi:1976iz,Ebert:1982pk,Volkov:1984kq,Volkov:1986zb,Ebert:1985kz,Vogl:1991qt,Klevansky:1992qe,Volkov:1993jw,Ebert:1994mf,Volkov:1999yi,Buballa:2003qv,Volkov:2005kw}. In the extended version of the model, the first radially excited states of the pseudoscalar, vector and axial vector mesons are included by using the polynomial form factors depending of the transverse momentum of the quark at the level of the quark-meson Lagrangian obtained after bosonisation~\cite{Volkov:2017arr,Volkov:1996br,Volkov:1996fk,Volkov:1999yi}. At the same time, the values of the main model parameters and coupling constants of mesons with quarks are similar to the values in the standard NJL model. The model not only explains the spontaneous breaking of chiral symmetry and dynamic appearance of quark masses, but also successfully describes the mass spectrum of mesons and their interactions using a limited number of parameters~\cite{Volkov:2005kw,Volkov:2017arr}. It is important to note that the NJL model naturally includes the vector dominance hypothesis. At the same time, the description of interactions and decay widths does not require introduction of additional arbitrary parameters. This provides the predictive power of the model and a unified method for describing different processes.
 
In the present work, we study the decays of the excited vector mesons $\rho(1450)$, $\omega(1420)$, $\phi(1680)$ and $K^{*}(1410)$ with production of three pseudoscalar mesons with different combinations of final particles. The channels for direct production of final mesons and the channels with intermediate vector mesons are calculated. As a result, we present predictions of decay widths that are currently insufficiently studied experimentally.

\section{Lagrangian of the extended NJL model}
The fragment of the Lagrangian of the extended NJL model containing the vertices that we need takes the form~\cite{Volkov:2017arr,Volkov:1999yi,Volkov:2005kw,Volkov:2024plj}:
\begin{eqnarray}
	\label{Lagrangian}
		\Delta L_{int} & = &
		\bar{q} \biggl[ 
		i \gamma^{5} \sum_{j = \pm,0} \lambda_{j}^{\pi} \left(a_{\pi}{\pi}^{j} + b_{\pi}{\pi'}^{j}\right) + i \gamma^{5} \sum_{j = \pm,0} \lambda_{j}^{K} \left(a_{K}{K}^{j} + b_{K}{K'}^{j}\right)  \nonumber \\ 
        && 
        + i \gamma^{5} \lambda_{0}^{K} \left(a_{K}\bar{K}^{0} + b_{K}\bar{K'}^{0}\right)
        + \frac{1}{2} \gamma^{\mu} \sum_{j = \pm,0} \lambda_{j}^{K^*} \left(a_{K^*}{K^*}_{\mu}^{j} 
        + b_{K^*}{K^{*}}'^{j}_{\mu}\right) 
        \nonumber\\
        && 
        + \frac{1}{2} \gamma^{\mu} \lambda_{0}^{K^*} \left(a_{K^*}\bar{K}_{\mu}^{*0} + b_{K^*}\bar{K}_{\mu}^{*'0}\right)
        + \frac{1}{2} \gamma^{\mu} \sum_{j = \pm,0} \lambda_{j}^{\rho} \left(a_{\rho}\rho^{j}_{\mu} + b_{\rho}\rho'^{j}_{\mu} \right)
        \nonumber\\
        && 
        + \frac{1}{2} \gamma^{\mu} \lambda^{\omega} \left(a_{\rho}\omega_{\mu} + b_{\rho}\omega'_{\mu} \right) + \frac{1}{2} \gamma^{\mu} \lambda^{\phi} \left(a_{\phi}\phi_{\mu} + b_{\phi}\phi'_{\mu} \right)	
		+ i\gamma_{5} \sum_{i = u, s} \lambda_{i} a^{i}_{\eta}\eta
		\biggl]q,
\end{eqnarray}
where $q$ and $\bar{q}$ are the triplets of u, d and s quarks, $\lambda$ are the linear combinations of the Gell-Mann matrices, the coefficients $a_{M}$ and $b_{M}$ take the following form:
\begin{eqnarray}
\label{coef}
	a_{M} = \frac{1}{\sin(2\theta_{M}^{0})}\left[g_{M}\sin(\theta_{M} + \theta_{M}^{0}) +
	g'_{M}f_{M}(k_{\perp}^{2})\sin(\theta_{M} - \theta_{M}^{0})\right], \nonumber\\
	b_{M} = \frac{-1}{\sin(2\theta_{M}^{0})}\left[g_{M}\cos(\theta_{M} + \theta_{M}^{0}) +
	g'_{M}f_{M}(k_{\perp}^{2})\cos(\theta_{M} - \theta_{M}^{0})\right],
\end{eqnarray}
where $M$ designates an appropriate meson. 

Since in the case of the $\eta$ meson four states are mixed, the appropriate factors have a different structure~\cite{Volkov:2017arr,Volkov:2024plj}:
\begin{eqnarray}
    a^{u}_{\eta} & = & 0.71 g_{\eta^{u}} + 0.11 g'_{\eta^{u}} f_{uu}(k_{\perp}^{2}), \nonumber\\
    a^{s}_{\eta} & = & 0.62 g_{\eta^{s}} + 0.06 g'_{\eta^{s}} f_{ss}(k_{\perp}^{2}).
\end{eqnarray}

The mixing angels of the ground and first radially excited meson states are presented in Table~\ref{tab_mixing}. They were fixed when building the model by using the meson masses.
\begin{table}[h!]
\begin{center}
\begin{tabular}{ccccccc}
\hline
   & $\pi$ & $K$ & $K^*$ & $\rho$ & $\omega$ & $\phi$ \\
\hline
$\theta_M$ & $59.48^{\circ}$	& $58.11^{\circ}$ & $84.74^{\circ}$ &  $81.80^{\circ}$  & $81.80^{\circ}$ & $68.4^{\circ}$  \\
$\theta^0_M$ & $59.12^{\circ}$ & $55.52^{\circ}$ & $59.56^{\circ}$ & $61.50^{\circ}$  & $61.50^{\circ}$ & $57.13^{\circ}$  \\
\hline
\end{tabular}
\end{center}
\caption{The values of the mixing angles~\cite{Volkov:2017arr,Volkov:1999yi,Volkov:2005kw}.}
\label{tab_mixing}
\end{table}

The form factor $f(k_\perp^2) = 1+dk_\perp^2$ was introduced to describe the first radially excited meson states. The constant $d$ is the slope parameter fixed based on the requirement that the form factor would not affect the vacuum condensate and, accordingly, would not change the constituent quark masses.

The coupling constants of mesons with quarks are the result of the redefinition of the meson fields which is necessary to give the kinetic term of the Lagrangian the standard form. Their fixing is ensured by calculating the one-loop quark contributions to the self-energy of mesons. Such constants of the quark-meson interaction take the following form~\cite{Volkov:2017arr,Volkov:1999yi,Volkov:2005kw}:
\begin{eqnarray}
 g_{\pi} = g_{\eta^{u}}=\left(\frac{4}{Z_{\pi}}I_{20}\right)^{-1/2}, &\quad&
 g'_{\pi}=g'_{\eta^{u}} =  \left(4 I_{20}^{f^{2}}\right)^{-1/2}, \nonumber\\
 g_{\eta^{s}}=\left(\frac{4}{Z_{\eta^s}}I_{02}\right)^{-1/2}, &\quad&
 g'_{\eta^{s}} =  \left(4 I_{02}^{f^{2}}\right)^{-1/2}, \nonumber\\
 g_{K} =\left(\frac{4}{Z_K}I_{11}\right)^{-1/2}, &\quad&
 g'_{K} =\left(4I^{f^2}_{11}\right)^{-1/2},  \\
 g_{K^*} =\left(\frac{2}{3}I_{11}\right)^{-1/2}, &\quad&
 g'_{K^*} =\left(\frac{2}{3}I_{11}^{f^{2}}\right)^{-1/2},  \nonumber\\
 g_{\rho} = g_{\omega} = \left(\frac{2}{3}I_{20}\right)^{-1/2}, &\quad&
 g'_{\rho} = g'_{\omega} = \left(\frac{2}{3}I_{20}^{f^{2}}\right)^{-1/2}, \nonumber\\
 g_{\phi} = \left(\frac{2}{3}I_{02}\right)^{-1/2}, &\quad&
 g'_{\phi} = \left(\frac{2}{3}I_{02}^{f^{2}}\right)^{-1/2},\nonumber
\end{eqnarray}
where $Z_{\pi}$, $Z_K$ and $Z_{\eta^{s}}$ are the additional renormalization constants appearing while taking into account the transitions between axial vector and pseudoscalar mesons:
\begin{eqnarray}
    Z_{\pi} = \left(1 - 6\frac{m_{u}^{2}}{M^{2}_{a_{1}}}\right)^{-1}, \quad Z_{K} = \left(1 - \frac{3}{2}\frac{\left(m_{u} + m_s\right)^{2}}{M^{2}_{K_{1A}}}\right)^{-1}, \quad  Z_{\eta^s} = \left(1 - 6\frac{m_{s}^{2}}{M^{2}_{f_{1}}}\right)^{-1},
\end{eqnarray}
where $m_u = m_d = 270$~MeV, $m_s = 420$~MeV are the constituent quark masses; $M_{a_{1}}$ and $M_{f_{1}}$ are the masses of the appropriate axial vector mesons; $M_{K_{1A}}$ is the reduced mass, the result of the mixing of the states $K_{1}(1270)$ and $K_{1}(1400)$:
\begin{eqnarray}
\label{MK1A}
    M^{2}_{K_{1A}} = \left(\frac{\sin^{2}{\alpha}}{M^{2}_{K_{1}(1270)}} + \frac{\cos^{2}{\alpha}}{M^{2}_{K_{1}(1400)}}\right)^{-1},
\end{eqnarray}
the mixing angle $\alpha = 57^{\circ}$~\cite{Volkov:2019awd}.

The integrals appearing in the quark loops while renormalization of the Lagrangian take the form
\begin{eqnarray}
\label{integral_1}
	I_{n_{1}n_{2}}^{f^{m}} =
	-i\frac{N_{c}}{(2\pi)^{4}}\int\frac{f^{m}(k^2_{\perp})}{(m_{u}^{2} - k^2)^{n_{1}}(m_{s}^{2} - k^2)^{n_{2}}}\Theta(\Lambda_{3}^{2} - k^2_{\perp})
	\mathrm{d}^{4}k,
\end{eqnarray}
where $\Lambda_3=1030$~MeV is the cut-off parameter~\cite{Volkov:2017arr}.

\section{\label{sec3}Decays of the mesons $\rho'$, $\omega'$ and $\phi'$}
\begin{figure*}[t]
 \centering
   \centering
   \begin{tikzpicture}
    \begin{feynman}
      \vertex (k) {\(\rho' \)};
      \vertex [dot, right=1.5cm of k] (a) {};
      \vertex [dot, above right=1.8cm of a] (c){}; 
      \vertex [dot, below right=1.8cm of a] (e){};
      \vertex [dot, right=2.4cm of a] (d) {};
      \vertex [right=1.4cm of c] (f) {\(K^+\)};
      \vertex [right=1.4cm of d] (g) {\(K^-\)};
      \vertex [right=1.4cm of e] (h) {\(\pi^-\)};
      \diagram* {
         (k) -- [double] (a),
         (a) -- [fermion] (c),
         (c) -- [fermion] (d),
         (d) -- [fermion] (e),
         (e) -- [fermion] (a),
         (c) -- [double] (f),
         (d) -- [double] (g),
         (e) -- [double] (h),
      };
     \end{feynman}
    \end{tikzpicture}
   \caption{The box diagram describing the direct decay $\rho' \to K^-K^+ \pi^-$.}
 \label{diagrambox}
\end{figure*}
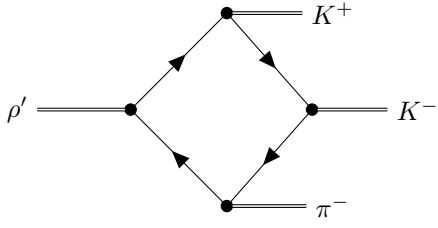%

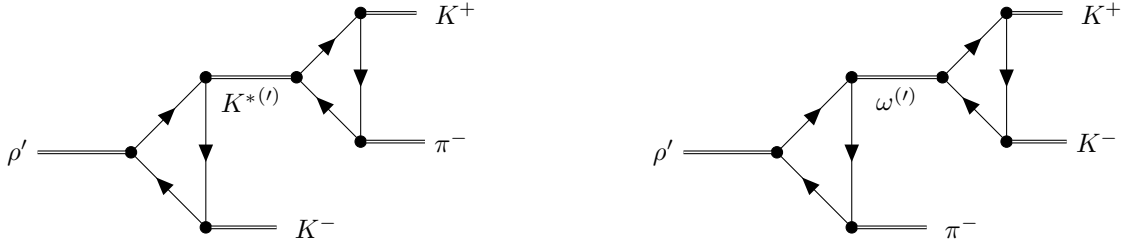
\begin{figure*}[t]
 \centering
  \begin{subfigure}{0.5\textwidth}
   \centering
   \begin{tikzpicture}
     \begin{feynman}
      \vertex (q) {\(\rho' \)};
      \vertex [dot, right=1.5cm of q] (d) {};      
      \vertex [dot, above right=1.4cm of d] (e) {};
      \vertex [dot, below right=1.4cm of d] (h) {};
      \vertex [dot, right=1.2cm of e] (f) {};
      \vertex [dot, above right=1.2cm of f] (n) {};  
      \vertex [dot, below right=1.2cm of f] (m) {};   
      \vertex [right=1.2cm of n] (l) {\(\ K^+ \)}; 
      \vertex [right=1.2cm of m] (s) {\(\pi^- \)};  
      \vertex [right=1.4cm of h] (k) {\(\ K^- \)}; 
      \diagram* {
         (q) -- [double] (a),         
         (d) -- [fermion] (e),  
         (e) -- [fermion] (h),
         (d) -- [anti fermion] (h),
         (e) -- [double, edge label'=\({{K^{*}}^{(\prime)}} \)] (f),
         (f) -- [fermion] (n),
         (n) -- [fermion] (m),
         (f) -- [anti fermion] (m), 
         (h) -- [double] (k),
         (n) -- [double] (l),
	 (m) -- [double] (s),
      };
     \end{feynman}
    \end{tikzpicture}
  \end{subfigure}%
 \centering
 \begin{subfigure}{0.5\textwidth}
  \centering
   \begin{tikzpicture}
     \begin{feynman}
      \vertex (q) {\(\rho' \)};
      \vertex [dot, right=1.5cm of q] (d) {};      
      \vertex [dot, above right=1.4cm of d] (e) {};
      \vertex [dot, below right=1.4cm of d] (h) {};
      \vertex [dot, right=1.2cm of e] (f) {};
      \vertex [dot, above right=1.2cm of f] (n) {};  
      \vertex [dot, below right=1.2cm of f] (m) {};   
      \vertex [right=1.2cm of n] (l) {\(\ K^{+} \)}; 
      \vertex [right=1.2cm of m] (s) {\( K^- \)};  
      \vertex [right=1.4cm of h] (k) {\(\ \pi^- \)}; 
      \diagram* {       
         (q) -- [double] (a),     
         (d) -- [fermion] (e),  
         (e) -- [fermion] (h),
         (d) -- [anti fermion] (h),
         (e) -- [double, edge label'=\({\omega^{(\prime)}} \)] (f),
         (f) -- [fermion] (n),
         (n) -- [fermion] (m),
         (f) -- [anti fermion] (m), 
         (h) -- [double] (k),
         (n) -- [double] (l),
	    (m) -- [double] (s),
      };
     \end{feynman}
    \end{tikzpicture}
  \end{subfigure}%
 \caption{The diagrams of the vector channels of the decay $\rho' \to K^+K^-\pi^-$. The vertices $\rho' K^-K^+\pi^-$ with two triangle quark loops connected with the virtual vector mesons $K^{*0}$ and $\omega$.}
 \label{diagram2}
\end{figure*}%

In the present section, we consider the decays of the vector mesons $\rho'$, $\omega'$ and $\phi'$ with production of $3P$ with various combinations of mesons in the final state, where $P=K, \pi, \eta$. We start with the decay $\rho'^- \to K^+K^-\pi^-$. The diagrams describing this decay are presented in Figs. \ref{diagrambox} and \ref{diagram2}.

The amplitude of this process in the NJL model takes the following form:
\begin{eqnarray}
    \mathcal{M}(\rho'^- \to K^+K^-\pi^-) = 4\left(\mathcal{M}_\omega + \mathcal{M}_{K^*} + \mathcal{M}_{\omega'} + \mathcal{M}_{K^{*'}} + \mathcal{M}_{box}\right) e_\mu^*(p) \epsilon^{\mu\nu\lambda\delta}p_{\pi\nu}p_{K^+\lambda}p_{K^-\delta},
\end{eqnarray}
where $e_\mu^*(p)$ is the polarization vector of the initial meson; $\epsilon^{\mu\nu\lambda\delta}$ is the absolutely antisymmetric tensor; $p_{\pi}$, $p_{K^+}$ and $p_{K^-}$ are the momenta of the final states; the terms in the brackets are the contributions from the diagrams with the intermediate mesons $\omega$, $K^{*0}$ and the box diagram, respectively:
\begin{eqnarray}
    \mathcal{M}_\omega & = & -4 m_u I_{30}^{\rho'\omega\pi} I_{11}^{\omega K K} Z_{\omega K K} BW_{\omega}(p_{K^+} + p_{K^-}),\nonumber\\
    \mathcal{M}_{K^*} & = & -2 m_u I_{c1}^{\rho'K^*K} I_{11}^{K^* K \pi} \left(Z_{K^* \pi K} + Z_{K^* K \pi}\right) BW_{K^*}(p_{K^+} + p_{\pi}), \nonumber\\
    \mathcal{M}_{\omega'} & = & -4 m_u I_{30}^{\rho'\omega'\pi} I_{11}^{\omega' K K} Z_{\omega' K K} BW_{\omega'}(p_{K^+} + p_{K^-}),\\
    \mathcal{M}_{K^{*'}} & = & -2 m_u I_{c1}^{\rho'K^{*'}K} I_{11}^{K^{*'} K \pi} \left(Z_{K^{*'} \pi K} + Z_{K^{*'} K \pi}\right) BW_{K^{*'}}(p_{K^+} + p_{\pi}), \nonumber\\
    \mathcal{M}_{box} & = & I_{c2}^{\rho'KK\pi}, \nonumber
\end{eqnarray}
where the intermediate meson is described with the Breit-Wigner propagator:
\begin{eqnarray}
    BW_{meson}(q) = \frac{1}{M^2_{meson} - q^2 - i\sqrt{q^2}\Gamma_{meson}}.
\end{eqnarray}
The combinations of the convergent integrals appearing in the vertices with the initial vector meson are
\begin{eqnarray}
\label{conv}
    I_{c1} & = & I_{21} + (m_s - m_u) m_u I_{31}, \nonumber\\
    I_{c2} & = & (m_s - 3 m_u) I_{31} - 3 (ms - mu) m_u^2 I_{41}.
\end{eqnarray}

The integrals appearing in the quark loops take the following form:
\begin{eqnarray}
	I_{n_{1}n_{2}}^{M_1 M_2\dots M'_1 M'_2\dots} =
	-i\frac{N_{c}}{(2\pi)^{4}}\int\frac{a_{M_1}a_{M_2}\dots b_{M_1}b_{M_2}\dots}{(m_{u}^{2} - k^2)^{n_{1}}(m_{s}^{2} - k^2)^{n_{2}}}\Theta(\Lambda_{3}^{2} - k^2_{\perp})
	\mathrm{d}^{4}k,
\end{eqnarray}
where $a_M$ and $b_M$ are defined in (\ref{coef}).
The coefficients $Z_{\omega K K}$, $Z_{K^* \pi K}$ and $Z_{K^* K \pi}$ have appeared while taking into account the transitions between axial vector and pseudoscalar mesons:
\begin{eqnarray}
\label{Z}
	Z_{\omega^{(')} K K} & = & 1 - \frac{I_{11}^{\omega^{(')} K_1 K} I_{11}^{K_1 K}}{I_{11}^{\omega^{(')} K K}} \frac{(m_s + m_u)^2}{M_{K_{1A}}^2}, \nonumber\\
    Z_{K^{*(')} \pi K} & = & 1 - 2 \frac{I_{11}^{K^{*(')}Ka_1}I_{20}^{a_1\pi}}{I_{11}^{K^{*(')} K \pi}}\frac{m_u (3 m_u - m_s)}{M_{a_1}^2},\\
    Z_{K^{*(')} K \pi} & = & 1 - 2 \frac{I_{11}^{K^{*(')}K_1\pi}I_{11}^{K_1K}}{I_{11}^{K^{*(')} K \pi}} \frac{m_s (m_s + m_u)}{M_{K_{1A}}^2}\nonumber,
\end{eqnarray}
where $M_{K_{1A}}$ is defined in (\ref{MK1A}).\\

The amplitude of the process $\rho'^0 \to \pi^+ \pi^- \eta$ in the extended NJL model takes the following form:
\begin{eqnarray}
    \mathcal{M}(\rho'^0 \to \pi^+ \pi^- \eta) = -8\left(\mathcal{M}_\rho +  + \mathcal{M}_{\rho'} + \mathcal{M}_{box}\right) e_\mu^*(p) \epsilon^{\mu\nu\lambda\delta}p_{\eta\nu}p_{\pi^+\lambda}p_{\pi^-\delta}.
\end{eqnarray}
The contributions from the diagrams with the intermediate meson $\rho^0$ and the box diagram are presented bellow:
\begin{eqnarray}
    \mathcal{M}_\rho & = & 4 m_u I_{30}^{\rho'\rho\eta} I_{20}^{\rho \pi \pi} Z_{\rho \pi \pi} BW_{\rho}(p_{\pi^+} + p_{\pi^-}),\nonumber\\
    \mathcal{M}_{\rho'} & = & 4 m_u I_{30}^{\rho'\rho'\eta} I_{20}^{\rho' \pi \pi} Z_{\rho' \pi \pi} BW_{\rho'}(p_{\pi^+} + p_{\pi^-}),\\
    \mathcal{M}_{box} & = & m_u I_{40}^{\rho'\pi\pi\eta}, \nonumber
\end{eqnarray}
where the constant $Z_{\rho \pi \pi}$ describes the $a_1-\pi$ transition:
\begin{eqnarray}
	Z_{\rho^{(')} \pi \pi} & = & 1 - 4\frac{I_{20}^{\rho^{(')} a_1 \pi} I_{20}^{a_1 \pi}}{I_{20}^{\rho^{(')} \pi \pi}} \frac{m_u^2}{M_{a_1}^2}.
\end{eqnarray}

The amplitude of the process $\rho' \to \pi^- \pi^0 \eta$ is similar to the amplitude of the process $\rho'^0 \to \pi^+ \pi^- \eta$ with the replacement $\rho^0$ by $\rho^-$ in the intermediate state.\\

The amplitude of the process $\omega' \to \pi^0 K^+ K^-$ in the extended NJL model takes the following form:
\begin{eqnarray}
    \mathcal{M}(\omega' \to \pi^0 K^+ K^-) & = & 4 \biggl[ \mathcal{M}_\rho + \mathcal{M}_{K^{*+}} + \mathcal{M}_{K^{*-}} + \mathcal{M}_{\rho'} + \mathcal{M}_{K^{*+'}} 
    \nonumber \\  
    &&
    + \mathcal{M}_{K^{*-'}} + \mathcal{M}_{box}\biggl]  
    e_\mu^*(p) \epsilon^{\mu\nu\lambda\delta}p_{\pi\nu}p_{K^+\lambda}p_{K^-\delta}.
\end{eqnarray}
The contributions from individual diagrams:
\begin{eqnarray}
    \mathcal{M}_\rho & = & -4 m_u I_{11}^{\rho K K} I_{30}^{\omega'\rho\pi} Z_{\rho K K} BW_\rho(p_{K^+} + p_{K^-}), \nonumber\\
    \mathcal{M}_{K^{*+}} & = & -m_u I_{c1}^{\omega' K^* K} I_{11}^{K^*K\pi} \left(Z_{K^* \pi K} + Z_{K^* K \pi}\right) BW_{K^{*+}}(p_{\pi} + p_{K^+}), \nonumber\\
    \mathcal{M}_{K^{*-}} & = & -m_u I_{c1}^{\omega' K^* K} I_{11}^{K^*K\pi} \left(Z_{K^* \pi K} + Z_{K^* K \pi}\right) BW_{K^{*-}}(p_{\pi} + p_{K^-}), \nonumber\\
    \mathcal{M}_{\rho'} & = & -4 m_u I_{11}^{\rho' K K} I_{30}^{\omega'\rho'\pi} Z_{\rho' K K} BW_{\rho'}(p_{K^+} + p_{K^-}), \\
    \mathcal{M}_{K^{*+'}} & = & -m_u I_{c1}^{\omega' K^{*'} K} I_{11}^{K^{*'}K\pi} \left(Z_{K^{*'} \pi K} + Z_{K^{*'} K \pi}\right) BW_{K^{*+'}}(p_{\pi} + p_{K^+}), \nonumber\\
    \mathcal{M}_{K^{*-'}} & = & -m_u I_{c1}^{\omega' K^{*'} K} I_{11}^{K^{*'}K\pi} \left(Z_{K^{*'} \pi K} + Z_{K^{*'} K \pi}\right) BW_{K^{*-'}}(p_{\pi} + p_{K^-}), \nonumber\\
    \mathcal{M}_{box} & = & I_{c2}^{\omega'KK\pi}, \nonumber
\end{eqnarray}
where the constant $Z_{\rho K K}$ describes the $K_1-K$ transitions:
\begin{eqnarray}
	Z_{\rho^{(')} K K} & = & 1 - \frac{I_{11}^{\rho^{(')} K_1 K} I_{11}^{K_1 K}}{I_{11}^{\rho^{(')} K K}} \frac{(m_s + m_u)^2}{M_{K_{1A}}^2}.
\end{eqnarray}
The constants $Z_{K^* \pi K}$ and $Z_{K^* K \pi}$ are defined in (\ref{Z}). The combinations of the convergent integrals $I_{c1}$ and $I_{c2}$ are defined in (\ref{conv}).\\

The amplitude of the process $\omega' \to \pi^+\pi^-\pi^0$ in the NJL model takes the following form:
\begin{eqnarray}
    \mathcal{M}(\omega' \to \pi^+\pi^-\pi^0) & = & -8\biggl[ \mathcal{M}_{\rho^0} + \mathcal{M}_{\rho^+} + \mathcal{M}_{\rho^-} + \mathcal{M}_{\rho'^0} 
     \nonumber \\  &&
    + \mathcal{M}_{\rho'^+} + \mathcal{M}_{\rho'^-}  + \mathcal{M}_{box}\biggl] 
    e_\mu^*(p) \epsilon^{\mu\nu\lambda\delta}p_{\pi^0\nu}p_{\pi^+\lambda}p_{\pi^-\delta},
\end{eqnarray}
where
\begin{eqnarray}
    \mathcal{M}_{\rho^0} & = & 4 mu I_{30}^{\omega'\rho\pi} I_{20}^{\rho\pi\pi} Z_{\rho\pi\pi} BW_{\rho^0}(p_{\pi^+} + p_{\pi^-}), \nonumber\\
    \mathcal{M}_{\rho^+} & = & 4 mu I_{30}^{\omega'\rho\pi} I_{20}^{\rho\pi\pi} Z_{\rho\pi\pi} BW_{\rho^+}(p_{\pi^+} + p_{\pi^0}), \nonumber\\
    \mathcal{M}_{\rho^-} & = & 4 mu I_{30}^{\omega'\rho\pi} I_{20}^{\rho\pi\pi} Z_{\rho\pi\pi} BW_{\rho^-}(p_{\pi^-} + p_{\pi^0}), \nonumber\\
    \mathcal{M}_{\rho'^0} & = & 4 mu I_{30}^{\omega'\rho'\pi} I_{20}^{\rho'\pi\pi} Z_{\rho'\pi\pi} BW_{\rho'^0}(p_{\pi^+} + p_{\pi^-}), \\
    \mathcal{M}_{\rho'^+} & = & 4 mu I_{30}^{\omega'\rho'\pi} I_{20}^{\rho'\pi\pi} Z_{\rho'\pi\pi} BW_{\rho'^+}(p_{\pi^+} + p_{\pi^0}), \nonumber\\
    \mathcal{M}_{\rho'^-} & = & 4 mu I_{30}^{\omega'\rho'\pi} I_{20}^{\rho'\pi\pi} Z_{\rho'\pi\pi} BW_{\rho'^-}(p_{\pi^-} + p_{\pi^0}), \nonumber\\
    \mathcal{M}_{box} & = & 3 m_u I_{40}^{\omega'\pi\pi\pi}. \nonumber
\end{eqnarray}
\\

The amplitude of the decay $\phi \to K^+ K^- \pi^0$ in the NJL model takes the following form:
\begin{eqnarray}
    \mathcal{M}(\phi \to K^+ K^- \pi^0) = 4 \sqrt{2} \left(\mathcal{M}_{K^{*+}} + \mathcal{M}_{K^{*-}} + \mathcal{M}_{K^{*'+}} + \mathcal{M}_{K^{*'-}} + \mathcal{M}_{box}\right) e_\mu^*(p) \epsilon^{\mu\nu\lambda\delta}p_{\pi^0\nu}p_{K^+\lambda}p_{K^-\delta},
\end{eqnarray}
where
\begin{eqnarray}
    \mathcal{M}_{K^{*+}} & = & m_s I_{c3}^{\phi'K^{*}K} I_{11}^{K^{*}K\pi} \left(Z_{K^* \pi K} + Z_{K^* K \pi}\right) BW_{K^{*+}}(p_{\pi^0} + p_{K^+}), \nonumber\\
    \mathcal{M}_{K^{*-}} & = & m_s I_{c3}^{\phi'K^{*}K} I_{11}^{K^{*}K\pi} \left(Z_{K^* \pi K} + Z_{K^* K \pi}\right) BW_{K^{*-}}(p_{\pi^0} + p_{K^-}), \nonumber\\
    \mathcal{M}_{K^{*'+}} & = & m_s I_{c3}^{\phi'K^{*'}K} I_{11}^{K^{*'}K\pi} \left(Z_{K^{*'} \pi K} + Z_{K^{*'} K \pi}\right) BW_{K^{*'+}}(p_{\pi^0} + p_{K^+}), \\
    \mathcal{M}_{K^{*'-}} & = & m_s I_{c3}^{\phi'K^{*'}K} I_{11}^{K^{*'}K\pi} \left(Z_{K^{*'} \pi K} + Z_{K^{*'} K \pi}\right) BW_{K^{*'-}}(p_{\pi^0} + p_{K^-}), \nonumber\\
    \mathcal{M}_{box} & = & m_s I_{c4}^{\phi'KK\pi}. \nonumber
\end{eqnarray}
The combinations of the convergent integrals here have the form:
\begin{eqnarray}
I_{c3} & = & I_{12} - (m_s - m_u) m_s I_{13}, \nonumber\\
I_{c4} & = & I_{22} - m_s (m_s - m_u) I_{23}.
\end{eqnarray}

The amplitude of the decay $\phi \to K^+ K^- \eta$ in the extended NJL model takes the following form:
\begin{eqnarray}
    \mathcal{M}(\phi \to K^+ K^- \eta) & = & 4 \biggl[  \mathcal{M}_{\phi} + \mathcal{M}_{K^{*+}} + \mathcal{M}_{K^{*-}} + \mathcal{M}_{\phi'} 
     \nonumber \\  &&
    + \mathcal{M}_{K^{*'+}} + \mathcal{M}_{K^{*'-}} + \mathcal{M}_{box}\biggl] 
    e_\mu^*(p) \epsilon^{\mu\nu\lambda\delta}p_{\eta\nu}p_{K^+\lambda}p_{K^-\delta},
\end{eqnarray}
where
\begin{eqnarray}
    \mathcal{M}_{\phi} & = & 8 m_s I_{03}^{\phi'\phi\eta^s} I_{11}^{\phi K K} Z_{\phi K K} BW_{\phi}(p_{K^+} + p_{K^-}), \nonumber\\
    \mathcal{M}_{K^{*+}} & = & \sqrt{2} m_s I_{c3}^{\phi'K^{*}K} \left(I_{11}^{K^{*}K\eta^u} + \sqrt{2}I_{11}^{K^{*}K\eta^s}\right)\left(1 + Z_{K^{*}K\eta}\right)BW_{K^{*+}}(p_{\eta} + p_{K^+}), \nonumber\\
    \mathcal{M}_{K^{*-}} & = & \sqrt{2} m_s I_{c3}^{\phi'K^{*}K} \left(I_{11}^{K^{*}K\eta^u} + \sqrt{2}I_{11}^{K^{*}K\eta^s}\right)\left(1 + Z_{K^{*}K\eta}\right)BW_{K^{*-}}(p_{\eta} + p_{K^-}), \nonumber\\
    \mathcal{M}_{\phi'} & = & 8 m_s I_{03}^{\phi'\phi'\eta^s} I_{11}^{\phi' K K} Z_{\phi' K K} BW_{\phi'}(p_{K^+} + p_{K^-}), \\
    \mathcal{M}_{K^{*'+}} & = & \sqrt{2} m_s I_{c3}^{\phi'K^{*'}K} \left(I_{11}^{K^{*'}K\eta^u} + \sqrt{2}I_{11}^{K^{*'}K\eta^s}\right)\left(1 + Z_{K^{*'}K\eta}\right)BW_{K^{*'+}}(p_{\eta} + p_{K^+}), \nonumber\\
    \mathcal{M}_{K^{*'-}} & = & \sqrt{2} m_s I_{c3}^{\phi'K^{*'}K} \left(I_{11}^{K^{*'}K\eta^u} + \sqrt{2}I_{11}^{K^{*'}K\eta^s}\right)\left(1 + Z_{K^{*'}K\eta}\right)BW_{K^{*'-}}(p_{\eta} + p_{K^-}), \nonumber\\
    \mathcal{M}_{box} & = & \sqrt{2} \left(m_s I_{c4}^{\phi'KK\eta^u} + \sqrt{2}I_{c5}^{\phi'KK\eta^s}\right)\nonumber
\end{eqnarray}

The constants appearing while taking into account the transitions between axial vector and pseudoscalar mesons are
\begin{eqnarray}
    Z_{\phi^{(')} K K} & = & 1 - \frac{I_{11}^{\phi^{(')} K_1 K} I_{11}^{K_1K}}{I_{11}^{\phi^{(')} K K}}\frac{(m_s + m_u)^2}{M_{K_{1A}}^2},\nonumber\\
    Z_{K^{*(')}K\eta} & = & 1 - 2\frac{m_s I_{11}^{K^{*(')}K_1\eta^u} + \sqrt{2}m_u I_{11}^{K^{*(')}K_1\eta^s}}{I_{11}^{K^{*(')}K\eta^u} + \sqrt{2}I_{11}^{K^{*(')}K\eta^s}}I_{11}^{K_1K}\frac{m_s + m_u}{M_{K_{1A}}^2}
\end{eqnarray}
The combination of the convergent integrals is
\begin{eqnarray}
    I_{c5} = (3m_s - m_u)I_{13} - 3(m_s - m_u)m_s^2 I_{14}.
\end{eqnarray}

\section{\label{sec4}Decays $K^{*'}$}
The amplitude of the decay $K^{*'} \to K^- \pi^0 \eta$ contains the contributions from the box diagram and the diagram with the intermediate vector mesons with the decays ${K^{*}}^{(\prime)} \to K \pi$ and ${K^{*}}^{(\prime)} \to K \eta$ in the ground and excited states. The full amplitude of the decay represents the sum of these diagrams
\begin{eqnarray}
    \mathcal{M}({K^*}^{\prime -} \to K^- \pi^0 \eta) = -4\left(
    \mathcal{M}_{K^{*}} + \mathcal{M}_{K^{*'}} + \mathcal{M}_{box}\right)e_\mu^*(p) \epsilon^{\mu\nu\lambda\delta}p_{K\nu}p_{\pi^0\lambda}p_{\eta\delta},
\end{eqnarray}
where individual contributions of the channels take the form
\begin{eqnarray}
    \mathcal{M}_{K^*} & = & 4\left(m_u I_{c1}^{{K^*}'K^*\eta^u} + m_s \sqrt{2}I_{c3}^{{K^*}'K^*\eta^s}\right) I_{11}^{K^*K\pi} \left(Z_{K^*K\pi} + Z_{K^*\pi K}\right)BW_{K^{*-}}(p_{\pi} + p_{K})  \nonumber\\
    && 
    + 4 m_u I_{c1}^{{K^*}'K^*\pi} \left(I_{11}^{K^{*}K\eta^u} + \sqrt{2}I_{11}^{K^{*}K\eta^s}\right) \left(Z_{K^*K\eta}+1\right) BW_{K^{*-}}(p_{\eta} + p_{K}),
\end{eqnarray}
\begin{eqnarray}
    \mathcal{M}_{{K^*}'} & = & 4\left(m_u I_{c1}^{{K^*}'{K^*}'\eta^u} + m_s \sqrt{2}I_{c3}^{{K^*}'{K^*}'\eta^s}\right) I_{11}^{{K^*}'K\pi} \left(Z_{{K^*}'K\pi} + Z_{{K^*}'\pi K}\right)BW_{{K^*}^{\prime-}}(p_{\pi} + p_{K})  \nonumber\\
    && 
    + 4 m_u I_{c1}^{{K^*}'{K^*}'\pi} \left(I_{11}^{{K^*}'K\eta^u} + \sqrt{2}I_{11}^{{K^*}'K\eta^s}\right) \left(Z_{{K^*}'K\eta}+1\right) BW_{{K^*}^{\prime-}}(p_{\eta} + p_{K}),
\end{eqnarray}
\begin{eqnarray}
    \mathcal{M}_{box} & = & -4 \sqrt{2} m_u \left(I_{22}^{{K^*}'K\pi\eta^s} + m_u (m_s-m_u) I_{23}^{{K^*}'K\pi\eta^s}\right). 
\end{eqnarray}
\\

The amplitude of the decay $K^{*'-} \to K^- \pi^+ \pi^-$ in the extended NJL model takes the following form:
\begin{eqnarray}
    \mathcal{M}(K^{*'-} \to K^- \pi^+ \pi^-) = -4\left(\mathcal{M}_{\rho} + \mathcal{M}_{K^{*}} + \mathcal{M}_{\rho'} + \mathcal{M}_{K^{*'}} + \mathcal{M}_{box}\right)e_\mu^*(p) \epsilon^{\mu\nu\lambda\delta}p_{K\nu}p_{\pi^+\lambda}p_{\pi^-\delta},
\end{eqnarray}
where
\begin{eqnarray}
    \mathcal{M}_{\rho} & = & 4 m_u I_{c1}^{K^{*'}K\rho} I_{20}^{\rho\pi\pi} Z_{\rho\pi\pi} BW_{\rho}(p_{\pi^+} + p_{\pi^-}), \nonumber\\
    \mathcal{M}_{K^{*}} & = & 2 m_u I_{c1}^{K^{*'}K^{*}\pi} I_{11}^{K^{*}K\pi} \left(Z_{K^{*}\pi K} + Z_{K^{*}K\pi}\right) BW_{K^{*}}(p_{K} + p_{\pi^+}), \nonumber\\
    \mathcal{M}_{\rho'} & = & 4 m_u I_{c1}^{K^{*'}K\rho'} I_{20}^{\rho'\pi\pi} Z_{\rho'\pi\pi} BW_{\rho'}(p_{\pi^+} + p_{\pi^-}), \\
    \mathcal{M}_{K^{*'}} & = & 2 m_u I_{c1}^{K^{*'}K^{*'}\pi} I_{11}^{K^{*'}K\pi} \left(Z_{K^{*'}\pi K} + Z_{K^{*'}K\pi}\right) BW_{K^{*'}}(p_{K} + p_{\pi^+}), \nonumber\\
    \mathcal{M}_{box} & = & - I_{c2}^{K^{*'}K\pi\pi}. \nonumber
\end{eqnarray}

The processes $K^{*'-} \to \bar{K}^0 \pi^- \pi^0$ and $K^{*'-} \to K^- 2\pi^0$ have a similar structure to the process $K^{*'-} \to K^- \pi^+ \pi^-$. The channel with the intermediate $\rho^-$ meson of the process $K^{*'-} \to \bar{K}^0 \pi^- \pi^0$ differs from the channel with the intermediate $\rho^0$ meson of the process $K^{*'-} \to K^- \pi^+ \pi^-$ by the additional factor $\sqrt{2}$. The channels with the intermediate mesons $\bar{K}^{*0}$ and $K^{*-}$ of the process $K^{*'-} \to \bar{K}^0 \pi^- \pi^0$ differ from the appropriate channel with the meson $\bar{K}^{*0}$ of the process $K^{*'-} \to K^- \pi^+ \pi^-$ by division by $\sqrt{2}$. The box diagram of the process $K^{*'-} \to \bar{K}^0 \pi^- \pi^0$ differs from th box diagram of the process $K^{*'-} \to K^- \pi^+ \pi^-$ by multiplying by $\sqrt{2}$. The process $K^{*'-} \to K^- 2\pi^0$ contains only the channel with the intermediate meson $K^{*-}$ that differs from the channel with the intermediate meson $\bar{K}^{*0}$ of the process $K^{*'-} \to K^- \pi^+ \pi^-$ by the factor $\frac{1}{2}$, and also by taking into account the identity of the pions.

\section{Numerical estimations}
The results of the calculations of the considered processes are presented in Table~\ref{table2}. In the second column, the contributions to the decay widths from the box diagrams describing the direct production of the final states are presented. The contributions from the vector channels are presented in the third column. The total values for all possible vector channels with the intermediate mesons $\rho$, $\rho'$, $\omega$, $\omega'$, $K^*$ and ${K^*}^{\prime}$ are presented in the fourth column. For example, the vector channel of the decay $\rho'\to KK\pi$ contains the contributions of the intermediate states $\omega$, $\omega'$, $K^*$ and ${K^*}'$. In the decay ${K^*}' \to K \pi\pi$ in the vector channel, there are intermediate states $\rho$, $\rho'$, $K^*$ and ${K^*}'$. For all the considered processes the amplitudes of the possible vector channels are defined in sections \ref{sec3} and \ref{sec4}. The fourth column of Table~\ref{table2} contains the final width of the decays taking into account all channels. Among the described decays, the width of the decay $\omega'\to3\pi$ is distinguished by a large value $\Gamma(\omega'\to3\pi) = 358.31$~MeV. The obtained theoretical estimation is in agreement with the experimental value of the full width $\Gamma^{exp.}_{\omega'} = 590 \pm 90$~MeV obtained while analyzing the cross section of the process $e^+e^- \to \pi^+\pi^-\pi^0$~\cite{Achasov:2024tfh,Dimova:2025dxn}.

\begin{table}[h!!]
\begin{center}
\begin{tabular}{ccccccc}
\hline
Decay   & Box diagram &Vector channel & Decay width  \\
\hline
$\rho' \to K^+K^-\pi^-$ & 0.24 & 2.76 & 3.31 \\
$\rho' \to \pi^+\pi^-\eta$ & 1.16 & 15.48 & 20.58 \\
$\omega' \to \pi^+\pi^-\pi^0$ & 60.06 & 330.13 & 358.31 \\
$\omega' \to K^+K^-\pi^0$ & 0.11 & 0.41 & 0.76 \\
$\phi' \to K^+K^-\pi^0$ & 1.85 & 82.78 & 82.18 \\
$\phi' \to K^+K^-\eta$ & 0.10 & 22.15 & 22.56 \\
${K^*}^{\prime} \to K^-\pi^0\eta$ & 0.0011 & 0.0225 & 0.0335 \\
${K^*}^{\prime} \to K^-\pi^-\pi^+$ & 9.16 & 24.4 & 31.09 \\
${K^*}^{\prime} \to K^0\pi^-\pi^0$ & 18.16 & 33.79 & 47.19 \\
${K^*}^{\prime} \to K^-\pi^0\pi^0$ & 0 & 3.46 & 3.46 \\
\hline
\end{tabular}
\end{center}
\caption{The widths of the three-particle pseudoscalar decays of the first radially excited vector mesons in MeV}
\label{table2}
\end{table}

The theoretical error of the model is estimated at the level of 15\%~\cite{Volkov:2005kw,Volkov:2017arr}. The main source affecting the model precision is the break of the chiral symmetry related mainly to the current quark masses that are not equal to zero. This error also includes the uncertainties of the meson masses. According to the statistical analysis of numerous calculations, in most cases, in the framework of the model one can achieve satisfactory agreement with experimental data at low energies within the error indicated above.

\section{\label{Concl}Conclusion}
In the current work, the widths of the three-particle pseudoscalar decays of the first radially excited vector mesons of the $U(3)$ nonet $\rho’$, $\omega’$, $\phi’$ and $K^{*\prime}$ have been calculated. The theoretical estimations for the widths of the decays are presented in Table~\ref{table2}. The results obtained in the current work complement the widths of the main two-particle decays of the excited vector mesons calculated early in the NJL model~\cite{Volkov:1999yi,Volkov:2023hju}. The two-particle strong decays of the excited mesons were considered in the literature. In the work~\cite{Gutsche:2008qq}, the calculations were carried out in the model with the effective chiral Lagrangian developed in the context of the Chiral Perturbation Theory. The nonrelativistic quark model was applied for calculating strong decays in the works~\cite{Barnes:1996ff, Barnes:2002mu}. In the work~\cite{Mengesha:2013xab}, the decays $\rho' \to \omega\pi$ and  $\omega' \to \rho\pi$ were described by using the Bethe-Salpeter equation. The widths of the two-particle decays of the excited vector mesons in the NJL model do not contradict these works~\cite{Volkov:2023hju}..

Unfortunately, at the present time, there are no reliable experimental data for the considered decays. That is why we can present only indirect comparison with the experimental data for the processes $e^+e^- \to KK\pi$ and $e^+e^- \to KK\eta$, where according to the experiment the cross section is mainly determined by the contribution of the decays $\phi’\to K^*K \to KK\pi$ and $\phi’ \to \phi\eta \to KK\eta$ \cite{Ivanov:2019crp,SND:2020qmb, BESIII:2022wxz}. The prediction of the NJL model for the mass $M_{\phi’} = 1682$~MeV is in agreement with the fitted value $M^{exp.}_{\phi’} = 1673 \pm 5$~MeV from the data on the $e^+ e^-$ annihilation~\cite{BESIII:2022wxz}. On the other hand, the widths of the decays are $\Gamma_{NJL}(\phi’ \to KK\pi) = 87$~MeV and $\Gamma_{NJL}(\phi’ \to KK\eta) = 24$~MeV which is in agreement with the full width $\Gamma_{\phi'}=172 \pm 8$~MeV \cite{BESIII:2022wxz}.

It should be noted that the processes of the meson production in $e^+e^-$ annihilation and $\tau$ decays can be described in the unified model in the framework of the hypothesis of the vector current conservation. In the extended NJL model, numerous processes of meson production in colliding electron-positron beams and hadron $\tau$ decays were described~\cite{Volkov:2017arr}. Let us notice that numerous calculations of the different processes with participation of the first radially excited mesons also showed satisfactory agreement of the theoretical estimations obtained earlier with the experimental data~\cite{Volkov:1999yi, Volkov:2017arr}. It allows one to hope for reliability of the obtained results. The theoretical estimations of the widths of the decays can be useful for future experiments at the lepton colliders (Belle II, BES III, Super $c$-$\tau$) and highly intensive hadron factories.

\section*{Acknowledgments}
This research has been funded by the Science Committee of the Ministry of Science and Higher Education of the Republic of Kazakhstan Grant No. BR34637266.

\end{document}